\documentclass[pra,showpacs,twocolumn,aps]{revtex4}
 
\usepackage[normalem]{ulem}
\usepackage{exscale}
\usepackage{graphicx}
\usepackage{amsmath}
\usepackage{latexsym}
\usepackage{amsfonts}
\usepackage{amssymb}
\usepackage{times}

\usepackage{epsfig,pstricks}
\usepackage{psfrag}
\definecolor{darkgreen}{rgb}{0,0.4,0}
\newcommand{\beq}{\begin{equation}}
\newcommand{\beqa}{\begin{eqnarray}}
\newcommand{\eeq}{\end{equation}}
\newcommand{\eeqa}{\end{eqnarray}}
\def\id{{\rm 1\kern-.22em l}}

\newcommand{\ket}[1]{\left|#1\right>}
\newcommand{\bra}[1]{\left<#1\right|}   

\newcommand{\nn}{\nonumber\\}

\newcommand{\f}[1]{\mbox{\boldmath$#1$}}

\newcommand{\bea}{\begin{eqnarray}}
\newcommand{\ea}{\end{eqnarray}}
\newcommand{\eea}{\end{eqnarray}}
\newcommand{\ord}{\,{\cal O}}

\begin{document}

\title{Monogamy of entanglement and improved mean-field ansatz for spin 
lattices}

\author{Andreas Osterloh} 
\email{andreas.osterloh@uni-due.de}
\author{Ralf Sch\"utzhold}

\email{ralf.schuetzhold@uni-due.de}

\affiliation{
Fakult\"at f\"ur Physik, Universit\"at Duisburg-Essen, 
Lotharstrasse 1, 47057 Duisburg, Germany}

\date{\today}

\begin{abstract}
We consider rather general spin-$1/2$ lattices with large coordination 
numbers $Z$.
Based on the monogamy of entanglement and other properties of the concurrence 
$C$, we derive rigorous bounds for the entanglement between neighboring 
spins, such as $C\leq1/\sqrt{Z}$, which show that $C$ decreases for large $Z$.
In addition, the concurrence $C$ measures the deviation from mean-field 
behavior and can only vanish if the mean-field ansatz yields an exact ground 
state of the Hamiltonian. 
Motivated by these findings, we propose an improved mean-field ansatz by 
adding entanglement.  
\end{abstract}


\maketitle

\section{Introduction}

Quantum information theory is not only interesting in view of quantum 
computers and quantum cryptography, but offers important insights into 
other branches of physics as well. 
For instance, a deeper understanding of entanglement -- which is one of the 
major differences between classical and quantum physics -- can help us to 
grasp the complexity of quantum many-body problems better. 
This strategy has already lead to very successful 
developments,
for example matrix-product states, which have been shown to efficiently 
approximate ground states of suitable low-dimensional lattice Hamiltonians. 
For a recent review, see \cite{MPS}.
Unfortunately, transferring this concept to higher dimensional lattices 
with a consequently larger coordination
number $Z$  is a non-trivial task. 
Besides tensor-network states
\cite{ShiDuanVidal06,Tagliacozzo09,PizornVerstraete13}, 
a step into this direction is the quantum 
de~Finetti theorem \cite{definetti,Finetti,KrausLewCirac13}.  
In one version, this theorem implies the following statement:
If a given state $\hat\rho^{(n)}$ of $n\gg 1$ qubits is invariant under 
permutation of any two of those qubits, then the reduced density matrix of 
two qubits $\hat\rho^{(2)}$ can be approximated by a separable 
(i.e., non-entangled) state plus $\ord(1/n)$ corrections.  
However, ground states of lattice Hamiltonians typically do not obey the full 
permutational invariance required for this theorem to hold 
(unless we have a fully connected lattice where all sites are neighbors).
In the following, we replace this full permutational invariance by a much 
smaller sub-group, the lattice isotropy, and derive a similar statement based 
on the monogamy of entanglement \cite{Coffman00,Osborne06} 
and certain properties of the concurrence 
\cite{Wootters98,Uhlmann00,Abouraddy01}.

\section{Spin lattice}

Let us consider a general regular, isotropic, and bi-partite lattice of 
spins $1/2$ (i.e., qubits)  
described by the Hamiltonian 
\bea
\label{Hamiltonian}
\hat H
=
\frac{1}{Z}\sum_{<\mu,\nu>}
\hat{\f{\sigma}}_{\mu}\cdot\f{J}\cdot\hat{\f{\sigma}}_{\nu} 
+\sum_{\mu}\f{B}\cdot\hat{\f{\sigma}}_{\mu} 
\,,
\ea
where 
$\hat{\f{\sigma}}_\mu=(\hat\sigma^x_\mu,\hat\sigma_\mu^y,\hat\sigma^z_\mu)$ 
are the usual Pauli matrices acting on the spin at the lattice site $\mu$
and $\f{B}=(B_x,B_y,B_z)$ denotes the local field while $\f{J}$ is 
a $3\times3$-matrix (tensor) describing the interactions between neighboring 
sites $\mu$ and $\nu$ (denoted by $\scriptstyle <\mu,\nu>$).
Finally, $Z$ is the coordination number (i.e., it counts the number of 
neighbors $\nu$ for each given lattice site $\mu$), and we consider the limit
of large $Z$.
The $1/Z$-scaling in front of the $\f{J}$-term is chosen such that the energy
per lattice site remains well defined in this limit $Z\to\infty$.

In general, obtaining the ground state of a Hamiltonian of the from 
(\ref{Hamiltonian}) can be rather complicated. 
Here, we shall exploit the properties of entanglement in order to understand 
the features of this ground state better.
Obviously, the knowledge of the reduced density matrices 
$\hat\rho_{<\mu\nu>}$ of neighboring spins $\mu,\nu$ suffices for 
calculating the ground state energy. 
The entanglement between these sites $\mu$ and $\nu$ is also completely 
determined by $\hat\rho_{<\mu\nu>}$ and can be measured by the concurrence 
$C[\hat\rho_{<\mu\nu>}]$. 
This quantity satisfies the monogamy of entanglement, i.e., the one-tangle 
$\tau_1(\hat\rho_\mu)=4\det(\hat\rho_\mu)$ 
of a given lattice site $\mu$ described by the on-site
reduced density matrix $\hat\rho_\mu$ yields an upper bound to its 
entanglement with all neighboring sites $\nu$ via 
\cite{Coffman00,Osborne06}
\bea
\label{monogamy}
\tau_1(\hat\rho_\mu)=4\det(\hat\rho_\mu)\geq
\sum_{\nu}C^2[\hat\rho_{<\mu\nu>}]
\,.
\ea
Assuming that the ground state obeys the same (discrete) symmetries as the 
underlying lattice, those matrices $\hat\rho_{<\mu\nu>}$ have the same form 
for all $\nu$.
Thus the sum over $\nu$ just gives a factor $Z$ and we get the upper bound 
for the concurrence 
\bea
\label{concurrence}
C[\hat\rho_{<\mu\nu>}]
\leq
\sqrt{\frac{\tau_1}{Z}}
\leq
\sqrt{\frac{1}{Z}}
\,,
\ea
where we have used $\tau_1\leq 1$ in the last step. 
As a result, in the limit of large coordination numbers, the entanglement 
between two spins is suppressed with $1/\sqrt{Z}$ or even stronger 
(see below).
The entanglement between next-to-nearest neighbors $C'$ 
can be bound via similar arguments, for example in a hyper-cubic lattice 
in $D$ dimensions (where $Z=2D$), we get $C'\leq1/\sqrt{2D(D-1)}$.  

\section{Ground-state energy}

As our next step, we exploit the high symmetry (degeneracy) in the 
decomposition space for the concurrence (as a quadratic polynomial), 
which facilitates the decomposition of every two-qubit density matrix
\cite{Wootters98,Uhlmann00}
\bea
\hat\rho_{<\mu\nu>}
=
\sum\limits_{I=1}^{4}p_I\ket{\Psi_{\mu\nu}^I}\bra{\Psi_{\mu\nu}^I}
\ea
into (at most) four pure states $\ket{\Psi_{\mu\nu}^I}$ with the 
corresponding probabilities $p_I$ such that all these states 
$\ket{\Psi_{\mu\nu}^I}$ have the same concurrence $C$.
Then the properties of the concurrence enable us to split each state 
$\ket{\Psi_{\mu\nu}^I}$ into a separable part and an orthogonal entangled part
\cite{Abouraddy01}
\bea
\label{Bell}
\ket{\Psi_{\mu\nu}^I}
=
\sqrt{1-C}\,\ket{\psi_{\mu}^I}\,\ket{\psi_{\nu}^I}
+
\sqrt{C}\,\hat U^I_\mu\hat U^I_\nu\ket{{\rm Bell}}_{\mu\nu}
\,,
\ea
where $\ket{{\rm Bell}}$ is one of the maximally entangled Bell states such 
as $\ket{{\rm Bell}}=
(\ket{\uparrow\uparrow}+\ket{\downarrow\downarrow})/\sqrt{2}=\ket{\Phi^+}$ 
while $\hat U^I_\mu$ and $\hat U^I_\nu$ are some local unitary operations 
which do not change the entanglement.
Combining all these results, we get the following estimate for the energy 
per lattice site 
\bea
\label{energy}
\frac{\langle\hat H\rangle}{N}
&=&
\sum\limits_{I=1}^{4}
\frac{p_I}{2}
\left[
\langle\hat{\f{\sigma}}_{\mu}^I\rangle
\cdot\f{J}\cdot
\langle\hat{\f{\sigma}}_{\nu}^I\rangle
+\f{B}\cdot
\left(\langle\hat{\f{\sigma}}_{\mu}^I\rangle
+\langle\hat{\f{\sigma}}_{\nu}^I\rangle\right)
\right]
\nn
&&
+\ord(\sqrt{C})
\,,
\ea
where $\langle\hat{\f{\sigma}}_{\mu}^I\rangle=
\bra{\psi_{\mu}^I}\hat{\f{\sigma}}_{\mu}\ket{\psi_{\mu}^I}$
denote local (mean-field) expectation values.
The magnitude of the $\ord(\sqrt{C})$ corrections can be bounded from above 
by $(||\f{J}||+2||\f{B}||)\sqrt{C}$ where $||\f{J}||$ and $||\f{B}||$ are 
suitable norms such as $||\f{J}||=\sum_{ij}|J_{ij}|$.

Consequently, in the limit of large $Z$ and therefore small $C$, we may 
estimate the ground state energy (per lattice site) by the variational 
mean-field ansatz 
\bea
\label{mean-field}
\ket{\Psi_{\rm mf}}=\bigotimes\limits_\mu\ket{\psi_{\mu}}
\,.
\ea
Inserting this mean-field ansatz and minimizing the energy thus yields an 
estimate for the exact ground state energy up to $\ord(\sqrt{C})$ corrections. 
If this variational procedure yields a unique solution 
$\ket{\psi_{\mu}}=\ket{\psi_0}$, the resulting state $\ket{\psi_0}$ provides
a good approximation to the local (on-site) properties of the exact ground 
state.  

\section{Ising model}

Let us study this general procedure by means of an explicit example, 
the quantum Ising model 
\bea
\label{Ising-model}
\hat H
=
-\frac{J}{Z}\sum_{<\mu,\nu>}\hat{\sigma}_{\mu}^x\hat{\sigma}_{\nu}^x 
-B\sum_{\mu}\hat{\sigma}_{\mu}^z 
\,.
\ea
Up to an irrelevant global phase, the mean-field ansatz~(\ref{mean-field}) 
can be parametrized via 
\bea
\label{parametrize}
\ket{\psi_{\mu}}
=
\cos\frac{\vartheta_{\mu}}{2}\,\ket{\uparrow}+
e^{i\varphi_{\mu}}\sin\frac{\vartheta_{\mu}}{2}\,\ket{\downarrow}
\,,
\ea
and, after insertion into the Hamiltonian, we get the mean-field energy per 
lattice site 
\bea
\label{mean-energy}
\frac{\langle\hat H\rangle_{\rm mf}}{N}
&=&
-\frac{1}{2}
[J\sin\vartheta_\mu\cos\varphi_\mu\sin\vartheta_\nu\cos\varphi_\nu
\nn
&&
+B(\cos\vartheta_\mu+\cos\vartheta_\nu)]
\,.
\ea
Accordingly, for $B>|J|$, we obtain a unique minimum at 
$\vartheta_\mu=\vartheta_\nu=0$ corresponding to the paramagnetic state 
$\ket{\Psi_{\rm mf}}=\ket{\uparrow\uparrow\uparrow\dots}$. 

As stated above, this mean-field ansatz $\ket{\psi_{\mu}}=\ket{\uparrow}$
provides a good approximation to the local properties of the exact ground 
state for large $Z$ and thus small $C$.
To make this statement more precise, let us consider the on-site reduced 
density matrix of the exact ground state, which can be cast into the most 
general form  
\bea
\label{most-general}
\hat\rho_{\mu}
=
\left(1-p\right)\ket{\uparrow}\bra{\uparrow}
+
p\ket{\downarrow}\bra{\downarrow}
+
\alpha\ket{\uparrow}\bra{\downarrow}
+
\alpha^*\ket{\downarrow}\bra{\uparrow}
.
\ea
By invoking symmetry arguments, one can even show that $\alpha$ must vanish
exactly in the paramagnetic state, but this is not necessary for our purposes.
Using the parametrization (\ref{parametrize}) for the states 
$\ket{\psi_{\mu}^I}$ and Taylor expanding Eq.~(\ref{energy}) for small 
$\vartheta_{\mu,\nu}^I$, we find 
\bea
\label{compare}
\frac{\langle\hat H\rangle}{N}
&\geq&
\frac{\langle\hat H\rangle_{\rm mf},0}{N}
+
(B-|J|)
\sum\limits_{I=1}^{4}
\frac{p_I}{4}
\left[(\vartheta_{\mu}^I)^2+(\vartheta_{\nu}^I)^2\right]
\nn
&&
+\ord(p_I[\vartheta_{\mu,\nu}^I]^4)
+\ord(\sqrt{C})
\,,
\ea
where $\langle\hat H\rangle_{\rm mf}/N=-B$ is the mean-field energy per 
lattice site~(\ref{mean-energy}). 
Obviously, the exact ground state energy $\langle\hat H\rangle/N$ in the above 
expression must not exceed that of the mean-field ansatz 
$\langle\hat H\rangle_{\rm mf}/N$, which yields the bound 
$p_I(\vartheta_{\mu,\nu}^I)^2\leq\ord(\sqrt{C})$. 
This implies that the probability $p$ in Eq.~(\ref{most-general}) scales with 
$p\leq\ord(\sqrt{C})$. 
Analogously, one can obtain the  bound $\alpha\leq\ord(\sqrt[4]{C})$ 
consistent with the properties of $\hat\rho_{\mu}$ such as 
$\det(\hat\rho_{\mu})\geq0$ or ${\rm Tr}\{\hat\rho_{\mu}^2\}\leq 1$. 

As a result, we find that the one-tangle $\tau_1(\hat\rho_\mu)$ is also 
suppressed by $\tau_1\leq\ord(\sqrt{C})$.
Together with our initial bound $C\leq Z^{-1/2}$ from (\ref{concurrence}), 
we thus get $\tau_1\leq\ord(Z^{-1/4})$.
However, inserting this estimate back into Eq.~(\ref{concurrence}), 
we obtain the improved scaling $C\leq\sqrt{\tau_1/Z}\leq\ord(Z^{-5/8})$.
Repeatedly iterating this procedure, the scaling exponents eventually 
converge to 
\bea
C\leq\ord(Z^{-2/3})
\,,\quad
\tau_1\leq\ord(Z^{-1/3})
\,.
\ea
On the other hand, the hierarchy of correlations derived in 
\cite{Equilibration,Quasi-particle}, for example, 
suggests that the one-tangle as well as all two-point 
correlations are suppressed by $1/Z$ in this situation.  

Since the maximum two-point correlation cannot be smaller  
than the concurrence $C$ \cite{Popp04}, 
this would imply an even stronger bound 
$C\leq\ord(Z^{-1})$, but -- to the best of our knowledge -- 
there is no rigorous proof, yet.
Of course, the concurrence could be even smaller (see below). 

\section{Improved mean-field ansatz}

Having found that the concurrence $C$ measures the deviation from the 
mean-field behavior, let us try to use this insight in order to improve 
the mean-field ansatz by adding entanglement. 
Inspired by Eq.~(\ref{Bell}), we start with the following ansatz for two 
sites 
\bea
\label{two-sites}
\ket{\Psi_{\mu\nu}}={\cal N}
\left(1+\hat{\f{\sigma}}_{\mu}\cdot\f{\xi}\cdot\hat{\f{\sigma}}_{\nu} 
\right)
\ket{\uparrow}_\mu\ket{\uparrow}_\nu
\,,
\ea
where $\f{\xi}$ acts as as entangling operation leading to a small but 
non-zero concurrence and ${\cal N}$ is the normalization. 
For the paramagnetic state 
$\ket{\Psi_{\rm mf}}=\ket{\uparrow\uparrow\uparrow\dots}$
of the Ising model, it is sufficient to keep only the relevant operators 
$\xi\sigma_\mu^x\sigma_\nu^x$ (or $\xi\sigma_\mu^-\sigma_\nu^-$).
Applying this procedure to the whole lattice yields the improved 
mean-field ansatz 
\footnote{Alternatively, one could use the unitary operator 
$\exp\{\xi_{\mu\nu}\sigma_\mu^-\sigma_\nu^--{\rm h.c.}\}$
as entangling operation, which gives the same result to first order in 
$\xi_{\mu\nu}$ but deviates in higher orders.  
This entangling operator 
$\exp\{\xi_{\mu\nu}\sigma_\mu^-\sigma_\nu^--{\rm h.c.}\}$
has the advantage that it is unitary -- but, as a drawback, it does not 
factorize as the operation in Eq.~(\ref{imf}).
Note that the above procedure is somewhat similar to the coupled-cluster 
ansatz used in quantum chemistry, for example.}
\bea
\label{imf}
\ket{\Psi}_{\rm imf}=
{\cal N}
\left(
\prod_{<\mu,\nu>}
\exp\left\{\xi\hat\sigma_\mu^x\hat\sigma_\nu^x\right\}
\right)
\bigotimes\limits_\mu
\ket{\uparrow}_\mu
\,,
\ea
where $\ket{\Psi_{\rm mf}}=\bigotimes_\mu\ket{\uparrow}_\mu=
\ket{\uparrow\uparrow\uparrow\dots}$ is the original mean-field ansatz 
(without entanglement). 
Here, we apply this entangling operation to nearest neighbors only, 
but this can be generalized easily to 
$\xi_{\mu\nu}\hat\sigma_\mu^x\hat\sigma_\nu^x$. 
Using the identity 
$\exp\left\{\xi\hat\sigma^x_\mu\hat\sigma^x_\nu\right\}
= \id\,\cosh\xi + \hat\sigma^x_\mu\hat\sigma^x_\nu\,\sinh\xi$, 
we get the single-site reduced density matrix 
\beq
\label{single-site}
\hat\rho_\mu 
= 
\frac{1}{2}
\left(
\id + \left(\frac{\cos(2\Im\xi)}{\cosh(2\Re\xi)}\right)^{Z/2}
\hat\sigma_\mu^z
\right)
\,.
\eeq
The reduced density matrix for nearest neighbors reads
\bea
\hat\rho_{<\mu\nu>}
&=&
\frac{1}{4}
\Big[
\id+
\left(
\hat\sigma^x_\mu\hat\sigma^x_\nu-
\chi^{2(Z-1)}\hat\sigma^y_\mu\hat\sigma^y_\nu 
\right)
\tanh2\Re\xi 
\nn
&&+
\chi^{2(Z-1)}\hat\sigma^z_\mu\hat\sigma^z_\nu+
\chi^{Z}\left(\hat\sigma^z_\mu+\hat\sigma^z_\nu\right) 
\nn
&&+
\omega^Z
\left(\hat\sigma^x_\mu\hat\sigma^y_\nu+\hat\sigma^y_\mu\hat\sigma^x_\nu\right) 
\Big]
\,,
\ea
where we have used the following abbreviations 
\bea
\chi=\frac{\cos(2\Im\xi)}{\cosh(2\Re\xi)}
\,,\quad
\omega=\frac{\sin(2\Im\xi)}{\cosh(2\Re\xi)}
\,,
\ea
containing the real $\Re\xi$ and imaginary part $\Im\xi$ of $\xi$. 

In order to test whether the ansatz~(\ref{imf}) is really an improvement, 
let us consider the energy which reads 
\bea
\frac{\langle\hat H\rangle_{\rm imf}}{N}
=
-
\frac{J}{2}\tanh(2\Re\xi)
-
B\,\left(\frac{\cos(2\Im\xi)}{\cosh(2\Re\xi)}\right)^Z
\,.
\ea
We see that adding entanglement -- i.e., increasing $\xi$ -- lowers the 
interaction energy $\propto J$ but increases the on-site term $\propto B$.  
Furthermore, we find that only the real part of $\xi$ can actually lower 
the energy, while the imaginary part always leads to an increase. 
The imaginary part of $\xi$ generates a unitary transformation $\hat U$ 
which cannot lower the energy $\langle\hat H\rangle_{\rm imf}$.
As another way to see this, one can apply this unitary transformation $\hat U$ 
to the Hamiltonian~(\ref{Ising-model}) instead of the state~(\ref{imf}).
Obviously, the interaction term $\propto J$ remains invariant under this 
unitary transformation and thus still yields a zero expectation value, 
while the expectation value of the local term $\propto B$ can only increase.
Consequently, we choose $\xi$ to be real such that the operation acting on 
the mean-field state in Eq.~(\ref{imf}) is non-unitary. 
(Thus the normalization $\cal N$.)

As also expected from stationary perturbation type arguments, 
the minimum energy is reached for a finite value 
\bea
\label{xi_min}
\xi_{\rm min}=\frac{J}{2BZ}+\ord(1/Z^2)
\,.
\ea
Consistent with the previous observations, the entangling strength $\xi$ 
decreases for large $Z$.
In addition, because the energy of the improved mean-field ansatz~(\ref{imf})
lies below the mean-field value, we know that the concurrence must be 
non-zero. 
Let us specify the relevant quantities for this example.
The one-tangle obtained from Eq.~(\ref{single-site}) reads 
\bea
\tau_1=1-\frac{1}{[\cosh(2\xi_{\rm min})]^{2Z}}=
\frac{J^2}{B^2Z}+\ord(1/Z^2)
\,.
\ea
As a result, the concurrence must be suppressed according to 
$C\leq J/(BZ)+\ord(1/Z^2)$ in view of (\ref{concurrence}).
To test this bound, let us calculate the concurrence of the state~(\ref{imf}).
Since the entangling strength $\xi$ scales with $1/Z$ according to 
Eq.~(\ref{xi_min}), we introduce the scaling variable $\zeta=Z|\xi|$.
Then, an expansion into powers of $1/Z$ (for fixed $\zeta$) yields the 
concurrence 
\bea
\label{C_xi}
C=2\,\frac{\zeta-\zeta^2}{Z}\,\Theta(1-\zeta)+\ord(1/Z^2)
\,.
\ea
The positive contribution $+2\zeta/Z$ is basically the concurrence of the 
pure state~(\ref{two-sites}), up to $\ord(1/Z^2)$ corrections. 
The negative contribution $-2\zeta^2/Z$, on the other hand, stems from the 
fact that $\hat\rho_{<\mu\nu>}$ is a mixed state due to the entanglement with 
all the other neighboring sites $\lambda\neq\mu,\nu$ which are averaged over 
when obtaining $\hat\rho_{<\mu\nu>}$.

Thus, for small $\zeta\ll1$, the concurrence $C$ approximately saturates 
the bound $C\leq J/(BZ)+\ord(1/Z^2)$ from (\ref{concurrence}).
For larger $\zeta$, on the other hand, the concurrence $C$ lies below this 
bound, and for $\zeta\geq1$, it even vanishes -- as indicated by the Heaviside 
step function $\Theta(1-\zeta)$. 
For a vanishing concurrence $C=0$, the arguments above imply that the 
ansatz~(\ref{imf}) cannot yield an improvement over the usual mean-field 
ansatz~(\ref{mean-field}).
However, this does not lead to any inconsistency because this case 
$\zeta\geq1$ corresponds to $|\xi|\geq1/Z$ and therefore $|J|\geq2B$, which 
lies already far beyond the (mean-field) critical point at $B=|J|$. 
Moreover, the concurrence~(\ref{C_xi}) assumes its maximum at $\zeta=1/2$,  
which precisely coincides with this critical point.
Thus, increasing the entangling strength $\xi$ beyond this point does not 
result in a growing concurrence $C$ anymore. 
Quite intuitively, since $C$ measures the ability to gain energy compared 
to the mean-field ansatz, we do not obtain any further improvement beyond 
this point. 
Whether this interesting observation -- i.e., that the maximum concurrence 
coincides with the (mean-field) critical point -- is just accidental or a 
more general property should be the subject of further investigations. 

\section{Degenerate ground states}

Note that the above arguments require a unique mean-field solution 
$\ket{\psi_0}$. 
Let us briefly discuss the cases where this solution is not unique. 
For $J>|B|$, we are in the symmetry-breaking ferromagnetic regime 
(on the mean-field level) where the mean-field energy~(\ref{mean-energy}) 
has two minima -- one at 
$\varphi_\mu=\varphi_\nu=0$ 
and the other one at 
$\varphi_\mu=\varphi_\nu=\pi$. 
For $J\gg|B|$, these two minima move to 
$\vartheta_{\mu}=\vartheta_{\nu}=\pi/2$
corresponding to the states 
$\ket{\Psi^-_{\rm mf}}=\ket{\leftarrow\leftarrow\leftarrow\dots}$
and 
$\ket{\Psi^+_{\rm mf}}=\ket{\rightarrow\rightarrow\rightarrow\dots}$. 
Even though the mean-field solution $\ket{\psi_0}$ is not unique in this case, 
we might select one of the two states as our starting point and carefully proceed in the 
same way as before. 
The incoherent average of the results in the two cases corresponds to 
the mixed state $\hat\rho=(\ket{\Psi^+_{\rm mf}}\bra{\Psi^+_{\rm mf}}+
\ket{\Psi^-_{\rm mf}}\bra{\Psi^-_{\rm mf}})/2$.

The remaining region of parameter space $J<-|B|$ corresponds to the 
anti-ferromagnetic regime (again on the mean-field level) which also 
breaks the symmetry.
In a bi-partite lattice, where we do not have to deal with frustration,
we could again choose one of the two states as mean-field background and 
apply the same procedure. 
In the case of frustration, however, things become more complicated and 
we cannot find a consistent mean-field background.
In this case, the one-tangle could well be of order one and thus the 
concurrence could be much larger, possibly $C=\ord(1/\sqrt{Z})$. 

Finally, at the critical points $J=\pm B$, we do find a consistent 
mean-field background, but the estimates after Eq.~(\ref{compare}) 
do not apply anymore, and thus the one-tangle and 
concurrence could also be larger than they are well inside the 
paramagnetic phase, for example. 

\section{Conclusions \& Outlook}


We have considered a general regular isotropic spin lattice with local 
(on-site) terms and nearest-neighbor interactions of Ising
type~(\ref{Hamiltonian}). 
Assuming that the ground state shares the isotropy of the lattice
monogamy of entanglement~(\ref{monogamy}) implies that the concurrence $C$ 
between neighboring spins decreases at least as $C\leq1/\sqrt{Z}$ for large 
coordination numbers $Z$.  
Under certain assumptions (such as a unique mean-field minimum), the bound 
can be improved to $C\leq\ord(Z^{-2/3})$.  
On the other hand, unless the mean-field ansatz~(\ref{mean-field}) yields 
an exact ground state 
of the Hamiltonian (see also \cite{Kurmann,Illuminati-FF09}), 
the concurrence $C$ is non-zero for nearest neighbors.  
In addition, the difference between the exact ground state energy per lattice 
site and that of the mean-field ansatz is bounded by $\ord(\sqrt{C})$,
i.e., the nearest neighbor entanglement $C$ serves as a measure for the 
deviation from the mean-field solution. 

Motivated by these findings, we propose an improved mean-field
ansatz~(\ref{imf}) by adding a small amount of entanglement.
For the Ising model~(\ref{Ising-model}) in the paramagnetic regime, we show 
that this ansatz~(\ref{imf}) does indeed yield a better approximation to the 
ground state and that the one-tangle and the concurrence (\ref{C_xi}) scale 
with $1/Z$ in this case, consistent with \cite{Equilibration,Quasi-particle}. 
Even though this is reminiscent of the quantum de~Finetti theorem 
\cite{Finetti}, where the corrections do also scale with the inverse of the 
number $n$ of involved qubits, we would like to stress that the 
scaling~(\ref{C_xi}) is obtained in a different way (e.g., without assuming 
full permutational invariance).

For further improvements, it would be very desirable to study and extend the 
various properties of the concurrence (such as the monogamy of entanglement)
to other entanglement measures.
For example, instead of considering only bi-partite entanglement
(which can be measured by the concurrence), it would be very interesting to 
study e.g. tri-partite entanglement. 
Unfortunately, however, our understanding of these matters is still far 
from complete.

\section*{Acknowledgements}

R.S.\ acknowledges fruitful discussions with I.~Cirac, G.~Vidal, 
U.~Schollw\"ock, and many others. 
This work was supported by the DFG (SFB-TR12). 




\end{document}